\documentstyle{mn}

\outer\def\gtae {$\buildrel {\lower3pt\hbox{$>$}} \over 
{\lower2pt\hbox{$\sim$}} $}
\outer\def\ltae {$\buildrel {\lower3pt\hbox{$<$}} \over 
{\lower2pt\hbox{$\sim$}} $}
\def\rchi{{${\chi}_{\nu}^{2}$}}
\newcommand{\Msun} {$M_{\odot}$}
\newcommand{\Mwd} {$M_{wd}$}

\title[The mass of the accreting white dwarf in magnetic CVs]
{Determining the mass of the accreting white dwarf in magnetic CVs using
{\sl RXTE} data}
\author[G. Ramsay]
{Gavin Ramsay\\
Mullard Space Science Laboratory, University College London,
Holmbury St.Mary, Dorking, Surrey, RH5 6NT
}

\date{Accepted 20 Dec 1999:}

\begin{document}

\maketitle

\begin{abstract} 

We have extracted spectra of 20 magnetic Cataclysmic Variables (mCVs)
from the {\sl RXTE} archive and fitted them using the X-ray continuum
method of Cropper et al to determine the mass of the accreting white
dwarf in each system. We find evidence that the mass distribution of
these mCVs is significantly different to that of non-magnetic isolated
white dwarfs, with the white dwarfs in mCVs being biased towards
higher masses. It is unclear if this effect is due to a selection
effect or whether this reflects an real difference in the parent
populations.

\end{abstract}


\vspace{2.5cm}

\section{Introduction}

The mass of the accreting white dwarf in a magnetic cataclysmic
variable (mCV) is a fundamental property of these binary systems. It
is one of the main parameters characterizing the emission from the
accretion region. This is because the temperature at the shock front
($T>10^{8}$ K) is determined by the mass of the white dwarf,
$M_{wd}$. The post-shock region (the region between the shock front
and the surface of the white dwarf) cools mainly by bremsstrahlung
radiation. If the magnetic field is sufficiently strong,
($B$\gtae10MG), then a significant amount of energy can also be
radiated away as cyclotron radiation.

Recently several groups have made estimates of $M_{wd}$ in a number of
mCVs using X-ray data. These groups have used one of two techniques to
derive $M_{wd}$: the continuum method, where the slope of the X-ray
continuum is fitted, and the X-ray line method where the intensity
ratio of different emission line species is used. Both of these
methods have their difficulties. In the case of the line method
(Fujimoto \& Ishida 1997, Ezuka \& Ishida 1999), this method is more
suited in determining $M_{wd}$ for low mass systems. This is because
in high mass systems the lines that are used are formed close to the
surface of the white dwarf and therefore do not accurately reflect the
shock temperature (Cropper, Wu \& Ramsay 1999). For low mass systems
this is less of a problem. In the case of the continuum method
(Cropper, Ramsay \& Wu 1998 and Cropper et al 1999), absorption
effects can be complex and lead to poor fits to the data. For low mass
systems, these methods agree remarkably well: in the case of EX Hya
the best fit masses agree to within 0.02 \Msun.

In this paper, the continuum method is used to determine $M_{wd}$ in
20 mCVs. In a series of papers (Cropper, Ramsay \& Wu 1998 and Cropper
et al 1999), we have added refinements to our model of the post-shock
region. These improvements include adding the effects of cyclotron
radiation as a source of cooling, allowing for the fact that the
post-shock region is multi-temperature rather than single temperature,
and adding a gravitational potential over the height of the post-shock
region. While these improvements to our model do not necessarily
improve the statistical quality of the fits to the X-ray data, they
are essential if accurate masses are to be derived.

As well as modelling the emission spectrum of the post shock region
accurately, the absorption, both internal and external to the binary
system, has to be taken into account. Cropper et al (1999) showed that
the model chosen to account for this absorption can effect the
resulting value of $M_{wd}$ to a small degree. For instance, they
found some evidence that a partial covering model gave lower masses
than an ionised absorber. In reality, the absorption model will be
much more complex than either of these models.

A more surprising result from Ramsay et al (1998) was that the mass of
the eclipsing mCV XY Ari using {\sl Ginga}, {\sl RXTE} and {\sl ASCA}
data was significantly different in the different detectors. However,
they found the mass determined using {\sl RXTE} data was consistent
with that determined from eclipse studies. Since {\sl RXTE} has a much
higher energy response than {\sl Ginga} and {\sl ASCA} it samples an
energy range which is closer to the shock temperature. Because of this
we expect that the masses derived using {\sl RXTE} data will be more
accurate than that of Cropper et al (1999). Therefore to increase our
sample of accurate $M_{wd}$ in mCVs, we have extracted data from the
{\sl RXTE} archive. The resulting mass distribution will be compared
with that of non-magnetic white dwarfs.

\section{The Observations}

{\sl RXTE} was launched in 1995 Dec, its prime aim being to observe
sources with maximum time resolution and moderate energy resolution
($\sim$1keV at 6keV). From the {\sl RXTE} data archive we have
extracted data from 6 strong field mCVs (those systems where the
magnetic field is sufficiently strong -- $B\sim$10-200MG -- to
synchronize the spin period of the white dwarf with the binary orbit
-- the polars) and 13 low field field mCVs (those systems where the
magnetic field is not sufficiently strong -- $B\sim$1--10MG -- to
synchronise the spin and orbital periods -- the intermediate polars,
or IPs). It is not the intension to perform a complete analysis of the
data on each system: indeed, for several systems this has already been
done elsewhere. Rather, the objective is to extract suitable data from
each source and determine $M_{wd}$.

Data were extracted only if the full set of data was present in the
archive (rather than only satellite slewing data which is added to the
archive much earlier than the full set of data). Data were cleaned
using standard {\tt FTOOLS} procedures. To improve the signal-to-noise
ratio, we extracted data using only the top Xenon layer of each PCA.
Since {\sl RXTE} is not an imaging X-ray telescope, background
subtraction is a particularly important issue. The background is not
well characterised before 1996 April 15, so we did not extract data
taken before this date. The background was estimated using {\tt
PCABACKEST} v2.1b and we used the faint source models applicable to
the date of the observation.

Table \ref{obs} lists the sources for which data were extracted. It
also shows the date of the observations, the exposure of the spectrum
used in the analysis and the mean count rate (2--20keV) per PCU of
this spectrum. The total exposure of the observations was in most
cases greater than the exposure shown in Table \ref{obs}. In many mCVs
an X-ray spectral variation is observed over the course of the spin
period of the white dwarf and/or the binary orbital period in the case
of the IPs. In these systems, using an integrated spectrum may result
in either a poor fit to the data or may effect the resulting
determination of $M_{wd}$. Therefore, in the majority of systems a
spectrum was extracted which covered particular spin and/or orbital
phases: this is tabulated in the last column of Table \ref{obs}. In
addition, two of the polars in our sample are slightly asynchronous:
BY Cam and V1432 Aql. In order that a spectral variation due to the
spin-orbit beat interval was not present in the resulting spectrum, we
selected data covering a much shorter time interval than the beat
period (BY Cam: $P_{beat}$=14.5 days, V1432 Aql: $P_{beat}$=49.5
days).

\begin{table*}
\begin{center}
\begin{tabular}{lrrrr}
\hline
Source& Date & Exp (sec)& Ct/s/PCU&Phase\\
\hline
V1432 Aql& Jul 1996& 4176&3.3& Spin max\\
BY Cam& June 1997& 39712& 3.5& Orbital max\\
V834 Cen& Aug 1997& 17872&1.6&Integrated\\
AM Her& Aug 1998& 4048 & 16.7 & Spin max\\
BL Hyi& Sept 1997& 1584&3.9&Primary pole\\
V2301 Oph& May 1997& 5248&3.7&Spin max\\
CP Tuc& July 1997& 28496&2.8&Spin max\\
\hline
FO Aqr& May 1997& 25296&6.7&Spin max\\
XY Ari& Jul -- Aug 1996& 32384& 1.1& Integrated but excluding flare\\
V405 Aur& Apr 1996& 25568&2.8&Integrated\\
V709 Cas& Mar 1997& 13328&6.0& Spin min\\
BG CMi& Jan 1997& 44016&2.7& Integrated\\
TV Col& Aug 1996& 17440&6.6& Orbital max\\
TX Col& Mar 1997& 27600&2.4&Spin max\\
EX Hya& Jun 1996& 18144&9.6&Spin min (exclude orb min)\\
AO Psc& Sep 1997& 13888&6.6&Spin max (exclude orb min)\\
V1223 Sgr& Nov 1997& 6368&12.3&Spin max\\
V1062 Tau& Feb 1998& 22944&3.8&Spin max\\
RX J1238--38& Jan 1997& 29456&1.6&Spin max\\
RX J1712--24& May 1996& 7792& 13.8&Orbital max\\
\hline
\end{tabular}
\end{center}
\caption{The log of {\sl RXTE} observations of polars (top) and
intermediate polars (bottom) used in this paper. In columns 2 \& 3
we show the date of the observation and the exposure of the
spectrum used in the analysis. In the 4th column we show the mean
count rate (2--20keV) per PCU. In the 5th column we show which part of the
spin/orbital cycle has been used to make the spectrum used in the
analysis.}
\label{obs}
\end{table*}

\section{The Model}
\label{model}

In modelling the emission spectrum we use the multi-temperature model
of Cropper, Ramsay \& Wu (1998) with the modifications of Cropper et
al (1999) which includes the effects of varying the gravitational
potential within the height of the shock region. We take the mean
molecular mass of the plasma to be cosmic ($\mu$=0.615).  The ratio of
cyclotron to bremsstrahlung cooling, $\epsilon_{o}$, was fixed for
each system. In case of the IPs, this was fixed at the low value of
0.001. By inverting equation 10 of Wu et al (1994) we find we have
imposed a magnetic field strength of $B$=1--5 MG in these systems
(depending on the spectrum). For the polars, we fixed $\epsilon_{o}$
at a value which gave field strengths consistent with the known field
strength for each system.

For the absorption column, we use two basic models: one or more
partial covering models of cold material and the other a partially
ionised absorber of the type described by Cropper, Ramsay \& Wu
(1998). A cold absorber was also present in addition to both these
models to account for interstellar absorption. As noted above, both of
these models are expected to be simplifications of what is likely to
be present in mCVs.

\section{The fits}

A spectrum was extracted from each source covering the energy range
2--20keV and binned so there was a minimum of 50 counts in each
spectral element. We found that it was difficult to model the energy
range between 6.2--7.2keV in many systems. This is probably due to
reflection and fluorescence effects in this energy range which are
difficult to accurately model and also because there is an absorption
edge at 7.1keV which becomes prominent at high absorbing
columns. Because the Pulse Channel Analyser (PCA) has a limited energy
resolution ($\sim$1keV at 6keV), it is to be expected that excluding
some energy channels will effect the fit to some degree. To test this,
we used one of the highest signal to noise spectra (FO Aqr) and
excluded data in the energy range 6.0--7.5keV. This had no effect
($<$0.01\Msun) on the resulting best fit to $M_{wd}$ although the fit
was significantly improved. We therefore excluded this energy range in
our fits.

We show in Tables \ref{polfit} and \ref{ipfit} the best fit values of
$M_{wd}$ together with the goodness of fit for our sample of polars
and IPs. In the case of the polars we also show the fits obtained with
a single cold absorber. For the IPs a single cold absorber was not
sufficient to fit the data (with the exception of RX 1238--38) so
these fits are not shown in Table \ref{ipfit}. For two IPs, EX Hya and
RX J1712--24, and 2 polars, BY Cam and CP Tuc, good fits were not
obtained (the results of these poor fits are not shown in Tables
\ref{polfit} and \ref{ipfit}). 

There are several possible causes for these poor fits. One is poor
background subtraction. There is no evidence that there are problems
with background subtraction at the dates of these observations ({\sl
RXTE} helpdesk). Since the cosmic ray component of the background
model is dependent on galactic latitude, and the background model
assumes a high galactic latitude for the object in question, it is
possible that this could introduce an error, since the IPs with poor
fits have low galactic latitude (EX Hya: $b=+33^{\circ}$, RX J1712-24:
$b=+8^{\circ}$). However, other sources in our sample which have
similar, or even lower galactic latitude, have good fits. We can use
the fact that two sources in our sample are eclipsing systems (V2301
Oph and XY Ari) to test how well the background has been
subtracted. In both systems, the mean count rate during the eclipse is
consistent with zero.  Whilst we cannot completely exclude the
possibility that the poor fits are due to poor background subtraction,
it is more likely that the residuals are due to the effects of complex
absorption.

To test if it is possible to obtain good fits to the data by excluding
a larger energy range, we again examined the spectrum of FO Aqr in
more detail. Initially we excluded data below 8keV from the fitting
process. In comparison with the best fit mass obtained when we
included all the spectral channels except 6.0--7.5keV, we found the
best fit mass was within 0.05 \Msun of this best fit. However, the fit
was not very well constrained. By including the energy range covering
5-6keV (where there are no absorption edges) we found that the best
fit was within 0.05 \Msun and constrained to within 0.1 \Msun (at the
90 percent confidence interval). This gave us some confidence that
reasonably accurate masses could be obtained for EX Hya, RX J1712--24,
CP Tuc and BY Cam even when certain energy ranges are excluded. In the
case of EX Hya, RX J1712--24 and BY Cam, energies below 5keV and
between 6.0--7.5keV were excluded. In the case of CP Tuc, energies
between 6.0--8.5keV were excluded. The masses and goodness of fits to
the data shown in Table \ref{polfit} \& \ref{ipfit} for these 4
systems were obtained by these means.

\begin{table*}
\begin{center}
\begin{tabular}{lrrr}
\hline
Source& CA & CA + PC & CA + IA\\
      & \Msun (\rchi, dof, range)& \Msun (\rchi, dof, range) \Msun (\rchi, dof, range)\\
\hline
V1432 Aql& 1.34 (1.05, 47dof: $>$1.30)& 0.98 (0.56, 45dof:
0.78--1.19)& 1.27 (0.95, 45dof: $>$1.23)\\
BY Cam& 1.04 (1.04, 32dof: 0.95--1.16)& 1.04 (1.11, 30dof: 0.95--1.16)& 1.04 (1.12, 30dof: 0.95--1.16)\\
V834 Cen& 0.68 (0.74, 40dof, 0.63--0.72)& 0.66 (0.77, 38dof:
0.61--0.71)& 0.64 (0.71, 38dof: 0.62--0.71)\\
AM Her & 0.73 (0.98, 40dof: 0.70--0.76)& 0.74 (1.14, 38dof:
0.71--0.77)& 0.73 (1.03, 38dof: 0.70--0.76)\\ 
BL Hyi& 0.71 (0.55, 39dof: 0.63--0.78)& 0.71 (0.58, 37dof: 0.67--0.79)& 
0.71 (0.59, 37dof: 0.67--0.79)\\
V2301 Oph& 0.75 (0.95, 38dof: 0.70--0.80)& 0.74 (1.00, 36dof:
0.70--0.80)& 0.75 (0.99, 36dof: 0.70--0.80)\\
CP Tuc& 0.68 (1.37, 34dof: 0.65--0.70)& 0.73 (1.51, 32dof: 0.70--0.78) &
0.73 (1.51, 32dof: 0.70--0.78)\\
\hline
\end{tabular}
\end{center}
\caption{The results for the fits to our polar data. The fits using three
different models are shown. The emission model was the same in all
cases and is that described in Cropper et al (1999). In the first
model, the absorption is a cold absorber (CA), the next a CA plus
partial covering model (PC) and the third a CA plus ionised absorber
(IA). The energy range 6.0--7.5keV was excluded in the fits with the
exception of BY Cam and CP Tuc where an additional energy range was
excluded (see text). The
confidence interval for $M_{wd}$ is the 90 per cent interval. }
\label{polfit}
\end{table*}

\begin{table*}
\begin{center}
\begin{tabular}{lrr}
\hline
Source& CA + PC & CA + IA\\
      & \Msun (\rchi, dof, range)& \Msun (\rchi, dof, range)\\
\hline
FO Aqr& 0.88 (0.75, 35dof: 0.79--0.95)& poor fit\\
XY Ari& 0.97 (0.83, 46dof: 0.80--1.13)& 1.31 (1.13, 46dof: $>$1.25)\\
V405 Aur& 0.99 (1.09, 47dof: 0.88--1.14)& 1.32 (1.15, 49dof:
$>$1.28)\\
BG CMi& 1.15 (1.03, 44dof: 1.09--1.21)& 1.25 (0.97, 46dof:
1.20--1.29)\\
V709 Cas& 1.08 (1.22, 34dof: 0.91--1.13)& poor fit \\
TV Col& 0.97 (1.06, 37dof: 0.92--0.99)& 0.94 (1.02, 37dof:
0.88--0.97)\\
TX Col& 0.74 (0.72, 37dof: 0.69--0.79)& 0.72 (0.81, 37dof:
0.68--0.79)\\
EX Hya & 0.44 (1.05, 37dof: 0.41--0.47)& 0.47 (0.89, 37dof:
0.44-0.50)\\
AO Psc& 0.60 (1.10, 35dof: 0.57--0.63)& 0.61 (1.15, 35dof:
0.57--0.64)\\
V1223 Sgr& 1.07 (0.65, 45dof: 0.98--1.15)& 1.25 (0.95, 47dof:
1.21--1.27)\\
V1062 Tau& 0.86 (0.72, 45dof: 0.80--1.04)& 1.34 (0.75, 45dof:
$>$1.28)\\
RX J1238--38& 0.60 (1.00, 47dof: 0.57--0.66)& 0.60 
(1.01, 47dof: 0.57--0.66)\\
RX J1712--24& 0.71 (0.87, 36dof: 0.68--0.78)& 0.73 
(0.54, 36dof: 0.63--0.77)\\  
\hline
\end{tabular}
\end{center}
\caption{The results for the fits to our intermediate polar data. The table
follows the format of Table \ref{polfit}, with the exception of EX Hya
and RX J1712--24 where the energy range was smaller (see text for 
details).} 
\label{ipfit}
\end{table*} 

In the case of our polar sample, all systems, with the exception of
V1432 Aql, were well fitted with a simple cold absorber (the addition
of a more complex absorber did not improve the fit). This is not
altogether surprising since no disk is present in these systems. V1432
Aql is a near synchronous system, as is BY Cam where we had to ignore
data below 5keV to get a good fit. Observations of another near
synchronous system RX J2115--58 (Ramsay et al 1999) suggest that the
accretion flow in these systems is complex and at certain beat phases,
the stream does not attach itself onto the most favourable magnetic
field lines. This may increase the amount of absorption present in the
system and make it hard to model accurately.

In our sample of IPs, the fits using the cold absorber plus partial
covering model (CA+PC) was very similar to, or better than the model
using a cold absorber plus ionised absorber (CA+IA). In two systems
(FO Aqr and V709 Cas) good fits could not be achieved using the CA+IA
model. The mean mass for IPs with a CA+PC absorber model was lower
($M_{wd}=0.85\pm$0.21) compared with the CA+IA absorber model
($M_{wd}=0.96\pm$0.34). This is similar to that found by Cropper et al
(1999).

In the case of the eclipsing systems (the polar V2301 Oph and the IP
XY Ari) we can compare our results with that of mass estimates
obtained from eclipse studies. In the case of V2301 Oph we that find
our mass of 0.75 $\pm$0.05 \Msun is consistent with that of
0.80$\pm$0.06 \Msun (Silber et al 1994) and 0.9$\pm$0.1 \Msun (Barwig,
Ritter \& B\"{a}rnbantner (1994). In the case of XY Ari, Ramsay et al
(1998) found a mass of 0.78--1.03 \Msun. This compares with 0.80--1.13
\Msun using a CA+PC absorber model and $>$1.25 \Msun using CA+IA
absorber model. (We note that the masses quoted in Table \ref{ipfit}
for XY Ari differ from that quoted in Cropper et al (1999) since the
background model applied here is more accurate than that applied
previously). This together with the generally better fits obtained
using the CA+PC model for our IP sample suggests that this model is a
better approximation of the absorption present in mCVs compared with
the CA+IA model. We therefore use the masses obtained using this model
for the remainder of the paper.

\section{Discussion}

\subsection{White Dwarf masses derived using X-ray data}

We show in Table \ref{mass} the best fit masses for our sample derived
using {\sl RXTE} data along with the masses derived using {\sl Ginga}
data (taken from Cropper et al 1999) and {\sl ASCA} data (taken from
Ezuka \& Ishida 1999). The masses derived using {\sl Ginga} data were
obtained using the same continuum model as used here while those
derived using {\sl ASCA} data were determined using the X-ray line
analysis.

The mass estimates of Cropper et al (1999) using {\sl Ginga} data
were, on the whole, not very well constrained. Since {\sl RXTE} has a
higher energy response than {\sl Ginga} our new mass estimates are
much better constrained (the mean of the uncertainties in Tables
\ref{polfit} \& \ref{ipfit} is 0.15 \Msun). It is therefore not
surprising, that within the uncertainties, the {\sl RXTE} and {\sl
Ginga} masses using our continuum method are consistent, with the
exception of TX Col, which is heavier using the {\sl RXTE} data.

Ezuka \& Ishida (1999) determined the mass for 9 IPs using the
emission line technique. However, most of those masses were not very
well constrained. Only in the case of EX Hya (an uncertainty of 0.15
\Msun) and AO Psc (0.23 \Msun) were the uncertainties comparable with
that given here. In the case of EX Hya the masses derived from the
continuum and line methods are very similar (the best fit masses are
within 0.03 \Msun) and in the case of AO Psc, the mass derived using
the continuum method is marginally heavier than that of the line
method.  With a best fit mass of 0.45 \Msun, EX Hya is close to the
minimum mass for isolated C-O core white dwarfs (0.46 \Msun, Sweigart,
Greggio \& Renzini 1990). For masses lower than this, the white dwarf
is expected to be a He-core white dwarf formed as a result of mass
transfer in interacting binary stars.

\begin{table}
\begin{center}
\begin{tabular}{llrr}
\hline
Source& {\sl RXTE} & {\sl Ginga}& {\sl ASCA}\\
      &  \Msun & \Msun & \Msun \\
\hline
V1432 Aql& 0.98 & & \\
BY Cam& 1.04 & 0.98 & \\
V834 Cen& 0.68 & 0.54 & \\
AM Her& 0.73 & 0.85 & \\ 
BL Hyi& 0.71& & \\
V2301 Oph &0.75& & \\
CP Tuc& 0.68& & \\
RX J2115--58&0.79& & \\
\hline
FO Aqr& 0.88& 0.92 & 1.05\\
XY Ari& 0.97& 1.19& \\
V405 Aur& 1.10& & \\
V709 Cas&1.08& & \\
TV Col& 0.96& 1.30 & 0.51\\
TX Col& 0.73& 0.48& 0.66\\
BG CMi& 1.20& 1.09& \\
EX Hya& 0.45& 0.46& 0.48\\
AO Psc&0.60& 0.56& 0.40\\
V1223 Sgr&1.10& & 1.28\\
V1062 Tau& 0.90& & \\
RX J1238--38& 0.60& & \\
RX J1712--24& 0.71& & 0.68\\
\hline
\end{tabular}
\end{center}
\caption{The best fit mass of mCVs determined using {\sl RXTE} (this
paper except RX J2115--58 which was taken from Ramsay et al 1999). 
The mass determined using {\sl Ginga} data was taken from
Cropper et al (1999) and the {\sl ASCA} data from Ezuka \& Ishida
(1999).}
\label{mass}
\end{table}

Whilst we consider this sample of accreting magnetic white dwarfs
masses to be the best constrained sample to have been derived using
fits to X-ray data (using either the continuum or line method), a
level of caution is appropriate. Our model of the post-shock region
has undergone several improvements which make it more physically
realistic. These changes have resulted in slightly different mass
estimates. For instance, applying a gravitational potential across the
height of the post-shock region resulted in lower masses
($\sim$0.1\Msun) for masses above 1.0\Msun. While we believe that the
most important factors have been included in our model, other second
order effects have not. For instance, it is assumed that the stream is
approaching from infinity. In the case of IPs, which are thought to
have a truncated accretion disk of some sort, this is clearly not the
case. This will have a small effect on \Mwd. As noted above, the model
we use for the absorption is not appropriate and may have an effect on
\Mwd. In spite of this, we believe that the sample presented here is
the best available with which to compare masses of magnetic white
dwarfs found in accreting binary systems with that of non-magnetic
white dwarfs.

\subsection{The mass distribution of White Dwarfs mCVs}

In our sample of polars we find a mean mass of \Mwd=0.80$\pm$0.14
\Msun, while in our sample of IPs we find a mean mass of
\Mwd=0.85$\pm$0.21. Taken as a whole, we find a mean mass for our mCVs
of \Mwd=0.84$\pm$0.20 \Msun. A better test is the Kolmogorov-Smirnov
test (K-S test) which we can use to determine how likely it is that
two distributions come from the same parent population. We find that
using the K-S test there is a 81.4 per cent probability that our polar
and IP samples do not come from the same parent population. We do not
consider this to be significant.

\begin{figure}
\begin{center}
\setlength{\unitlength}{1cm}
\begin{picture}(8,6)
\put(-1.5,-2.5){\includegraphics{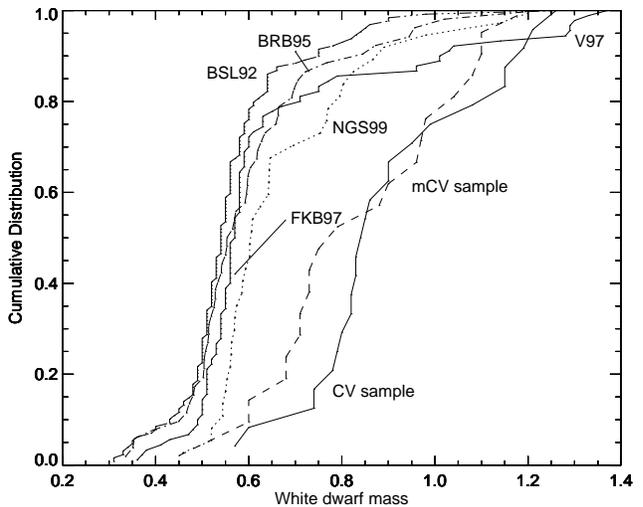}}
\end{picture}
\end{center}
\caption{The cumulative distributions for the mass of magnetic white
dwarfs in mCVs (this paper) and the mass of isolated non-magnetic white
dwarfs: Napiwotzki, Green \& Saffer (1999) (NGS99), Vennes et al (1997)
(V97), Bergeron, Saffer, Liebert (1992) (BSB92), Bragaglia,
Renzini \& Bergeron (1995) (BRL95), Finley, Koester, Basri (1997)
(FKB97). We also show the mass distribution of non-magnetic CVs taken
from Ritter \& Kolb (1998) (CV sample).}
\label{ks_plot} 
\end{figure}

To compare the combined distribution of our {\sl RXTE} mCV sample we
compare it with 5 samples of isolated non-magnetic white dwarf masses.
Vennes et al (1997) determined masses for 90 white dwarfs which were
detected in the {\sl EUVE} all-sky survey (hereafter V97). Finley,
Koester \& Basri (1997) (FKB97) determined the masses for 174 white
dwarfs which were taken from a sample of DA white dwarfs hotter than
$\sim$25000K. A total of 52 white dwarfs had mass determinations by
Bragaglia, Renzini \& Bergeron (1995) (BRB95) using a sample of bright
field white dwarfs. Bergeron, Saffer \& Liebert (1992) (BSL92)
determined the masses of 129 white dwarfs selected from the white
dwarf catalogue of McCook \& Sion (1987). Napiwotzki, Green \& Saffer
(1999) (NGS99) obtained masses for 43 white dwarfs selected from the
{\sl EUVE} and {\sl ROSAT} WFC all-sky surveys. White dwarfs which
were known to be in binary systems were excluded in our analysis. We
show their cumulative distributions along with our mCV sample in
Figure \ref{ks_plot}.

The most striking feature of Figure \ref{ks_plot} is that while all
the non-magnetic white dwarf distributions show a peak between
\Mwd$\sim$0.5--0.6\Msun, our mCV sample is much more uniformly
distributed. Performing a K-S test on the mCV distribution and the
non-magnetic isolated white dwarf distributions we find that they are
different with a greater than 99.99 percent probability. There is only
a 9.3$\times10^{-12}$ probability that the mCV distribution and the
distribution of BSL92 come from the same parent population.

We now compare the distribution of our {\sl RXTE} mCV sample with that
of non-magnetic CVs (nmCVs). For nmCVs the most reliable mass
determinations are those where the system is a double-lined
spectroscopic binary. There are 24 such systems in the catalogue of
Ritter \& Kolb (1998). We show the cumulative distribution of this
sample in Figure \ref{ks_plot}. Similar to our mCV distribution, the
nmCV distribution is also biased towards higher masses. The two
distributions are different at the 89 per cent level: we do not
consider this to be significant.

We now consider selection effects that can bias the mass distributions
which we have considered. All samples of white dwarf masses are biased
to some degree. For instance, FKB97 suggest that while studies based
on EUV selected white dwarfs will not preferentially select high mass
white dwarfs, white dwarfs with \Mwd$<$0.7 \Msun that are over 400pc
distant are strongly selected against since interstellar absorption
will effectively obscure them. Indeed, there are significant
differences between the various non-magnetic white dwarf mass
distributions. It is noteworthy that the EUV selected sample of V97
has a relatively high proportion of high mass systems: 12.2 percent of
its sample have masses over 1.0\Msun and 6.6 percent over
1.15\Msun. In the case of the nmCVs, a number of studies have looked
at selection effects in detail (eg Ritter \& Burkert 1986, Ritter et
al 1991, Politano 1996). For a magnitude limited sample, the sample is
biased towards more luminous objects which have a greater
gravitational potential and are therefore more massive. These studies
suggest that the apparent difference between the nmCV mass
distribution and the isolated white dwarf mass distribution can
largely be explained by this selection effect.

Selection effects are also important when considering our polar
sample, many of which are discovered in soft X-rays. Similar to EUV
selected white dwarfs, we may expect that low mass polars which are
more distant than a few hundred pc, are selected against. On the other
hand, we may expect that high mass polars are selected against for a
different reason. This is because in order that high mass white dwarfs
do not become unbound they must have a high internal magnetic pressure
(Suh \& Mathews 1999). Polars with high magnetic fields are not
included in the {\sl RXTE} database of polars because for high
magnetic fields a greater proportion of the emission from the post
shock region is emitted in the form of cyclotron radiation as opposed
to bremsstrahlung radiation and are thus not strong hard X-ray
emitters. Liebert (1988) found evidence for this in a sample of
isolated white dwarfs. Other massive non-accreting magnetic white
dwarfs have also recently been discovered: RX J0823.6--2525,
1.20$\pm$0.04 \Msun, 3MG (Ferrario, Vennes \& Wickramasinghe 1998), RE
J0317--858, 1.31--1.37 \Msun, $\sim$450MG (Ferrario et al
1997). 

While many polars have been discovered in all-sky soft X-ray surveys
such as that carried out using {\sl ROSAT}, IPs are much stronger in
hard X-rays. Apart from the {\sl HEAO-1} all-sky hard X-ray survey
which was undertaken in the late 1970's, no other such survey has been
made. Those IPs discovered in hard X-rays are likely, therefore, to
have been more luminous IPs and therefore more massive than average,
leading to a possible bias in their mass distribution.

This study has found evidence that the distribution of white dwarf
masses in mCVs is different from that of non-magnetic isolated white
dwarfs in the sense that there is a bias towards heavier masses for
white dwarfs in mCVs. On the other hand there is no significant
difference between the distribution of white dwarf masses in mCVs and
nmCVs. A detailed study of the various selections effects which are
relevant to mCVs is beyond the scope of this paper, but such a study
is needed to determine if (as in the case of nmCVs) selection effects
can account for the apparent difference between the distribution of
white dwarf masses in mCVs and isolated white dwarfs.

\section{Acknowledgments}

I would like to thank the referee, Boris G\"{a}nsicke, and Mark Cropper for
some useful comments on a previous draft of this paper.


\begin{thebibliography}{99} 

\bibitem{}Barwig, H., Ritter, H., B\"{a}rnbantner, O., 1994, A\&A,
288, 204
\bibitem{}Bergeron, P., Saffer, R. A., Liebert, J., 1992, ApJ, 394,
228 (BSL92)
\bibitem{}Bragaglia, A., Renzini, A., Bergeron, P., 1995, ApJ, 443,
735 (BRB95)
\bibitem{}Cropper, M., Ramsay, G., Wu, K., 1998, MNRAS, 293, 222
\bibitem{}Cropper, M., Wu, K., Ramsay, G., 1999, In 'Annapolis
Workshop on magnetic CVs', ASP Conf Ser, Vol 157, 325
\bibitem{}Cropper, M., Wu, K., Ramsay, G., Kocabiyik, A., 1999, 306, 809
\bibitem{}Ezuka, H., Ishida, M., ApJS, 120, 277
\bibitem{}Ferrario, L., Vennes, S., Wickramasinghe, D. T., Bailey,
J. A., Christian, D. J., 1997, MNRAS, 292, 205  
\bibitem{}Ferrario, L., Vennes, S., Wickramasinghe, D. T., 1998,
MNRAS, 299, L1
\bibitem{}Finley, D. S., Koester, D., Basri, G., 1997, ApJ, 488, 375 (FKB97)
\bibitem{}Fujimoto, R., Ishida, M., 1997, ApJ, 474, 774
\bibitem{}Liebert, J., 1988, PASP, 100, 1302
\bibitem{}McCook, G. P., Sion, E. M., 1987, ApJS, 65, 603 
\bibitem{}Napiwotzki, R., Green, P. J., Saffer, R. A., 1999, ApJ, 517,
399 (NGS99)
\bibitem{}Politano, M., 1996, ApJ, 465, 338
\bibitem{}Ramsay, G., Cropper, M., Hellier, C., Wu, K., 1998, MNRAS, 297, 1269
\bibitem{}Ramsay, G., Potter, S., Cropper, M., Buckley, D.,
Harrop-Allin, M. K., submitted, MNRAS
\bibitem{}Ritter, H., Burkert, A., 1986, A\&A, 158, 161
\bibitem{}Ritter, H., Politano, M., Livio, M., Webbink, R. F., 1991,
ApJ, 376, 177 
\bibitem{}Ritter, H., Kolb, U., 1998, A\&AS, 129, 83
\bibitem{}Silber, A. D., Remillard, R. A., Horne, K., Bradt, H. V.,
1994, ApJ, 424, 955
\bibitem{}Sweigart, A. V., Greggio, L., Renzini, A., 1990, ApJ, 364, 527
\bibitem{}Suh, I.-S., Mathews, G. J., submitted ApJ, astro-ph/9906239
\bibitem{}Wu, K., Chanmugam, G., Shaviv, G., 1994, ApJ, 426, 664
\bibitem{}Vennes, S., Thejll, P. A., Galvan, R. G., Dupuis, J., 1997,
ApJ, 480, 714 (V97)

\end{thebibliography}
\end{document}